\documentclass[11pt]{article}

\usepackage{amsmath,amssymb,amsthm,color,paralist,graphics,epsfig,graphicx,epstopdf}

\newtheorem{theorem}{Theorem}[section]
\newtheorem{assumption}[theorem]{Assumption}

\newtheorem{remark}[theorem]{Remark}
\newtheorem{lemma}[theorem]{Lemma}
\newtheorem{conjecture}[theorem]{Conjecture}

\begin{document}

%
%
%
%
%
%
%
%
%
%
%
%
%
%
%

\title{On a hypercycle equation with infinitely many members}

\author{Alexander S. Bratus$^{1,2,}$\footnote{e-mail: alexander.bratus@yandex.ru}$\,\,$, Olga~S.~Chmereva$^{3,} $\footnote{e-mail: o.s.ch@yandex.ru}\,, Ivan~Yegorov$^{4,} $\footnote{Also known as Ivan Egorov; e-mail: Ivan.Egorov@kla-tencor.com}\,, Artem S. Novozhilov$^{5,}$\footnote{Corresponding author; e-mail: artem.novozhilov@ndus.edu} \\[3mm]
\textit{\normalsize $^\textrm{\emph{1}}$Russian University of Transport, Obraztsova 15, Moscow 127994, Russia}\\[0mm]
\textit{\normalsize $^\textrm{\emph{2}}$Moscow Center of  Fundamental and  Applied Mathematics,}\\[-1mm]
\textit{\normalsize Lomonosov Moscow State University, Moscow 119992, Russia}\\[0mm]
\textit{\normalsize $^\textrm{\emph{3}}$Lomonosov Moscow State University, Leninskie Gory, MSU,}\\[-1mm]
\textit{\normalsize 2nd educational building, Moscow, 119991, Russia}\\[0mm]
\textit{\normalsize $^\textrm{\emph{4}}$KLA Corporation, 3 Technology Dr, Milpitas, CA 95035, USA}\\[0mm]
\textit{\normalsize $^\textrm{\emph{5}}$Department of Mathematics, North Dakota State University, Fargo, ND, 58108, USA}}

\date{}

\maketitle

\begin{abstract}A hypercycle equation with infinitely many types of macromolecules is formulated and studied both analytically and numerically. The resulting model is given by an integro-differential equation of the mixed type. Sufficient conditions for the existence, uniqueness, and non-negativity of solutions are formulated and proved. Analytical evidence is provided for the existence of non-uniform (with respect to the second variable) steady states. Finally, numerical simulations strongly indicate the existence of a stable nonlinear wave in the form of the wave train.


\end{abstract}


\section{Introduction}

An important class of replicator models involves systems of nonlinear ordinary differential equations with dynamics
restrained by the standard simplex in the state space and describes macromolecular interactions in various problems of
population genetics and evolutionary game theory \cite{SchusterSigmund1983,HofbauerSigmund1988,HofbauerSigmund1998},
as well as in theories of the origin of life \cite{Smith1979,EigenSchuster1982}.

Of special interest is the hypercycle model that was proposed by M.~Eigen and P.~Schuster \cite{EigenSchuster1982}.
It was related to the prebiotic evolution hypothesis stating that self-replicating molecules are predecessors of RNA, DNA and
eventually of cells. The hypercycle model has been thoroughly studied from the points of view of both population genetics
\cite{HofbauerSigmund1988} and mathematical frameworks for systems of nonlinear ordinary differential equations
\cite{SchusterSigmund1983,MalletParetSmith1990}.

\begin{figure}
\centering
\includegraphics[width=0.5\textwidth]{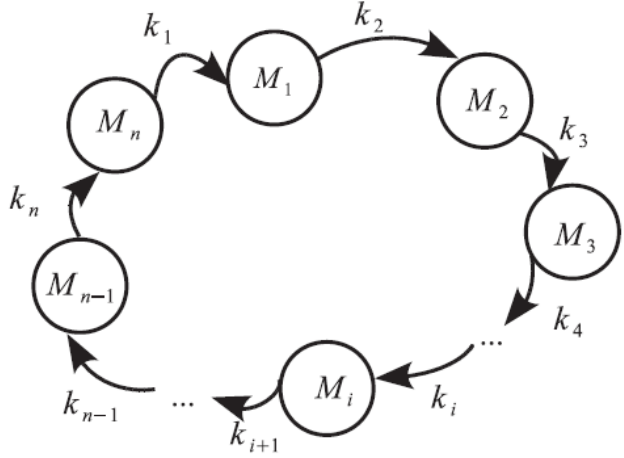}
\bf \caption{\rm A hypercyclic network.}
\label{Fig_1}
\end{figure}

A classical hypercycle is a finite closed network of self-replicating macromolecules (species) which are connected so that
each of them catalyzes the replication of the successor, with the last molecule reinforcing the first one; see Fig.~\ref{Fig_1}.
From the sociological perspective, the catalytic support for the replication of other molecules resembles altruistic behavior,
in contrast to conventional autocatalysis \cite{King1981,HofbauerSigmund1988}.

Let us briefly recall the mathematical formulation of the classical hypercycle model together with its key properties.

Consider a hypercycle with $ n $ macromolecules (species) labeled by $ \: i \, = \, 1, 2, \ldots, n \: $ (see
Fig.~\ref{Fig_1}), and denote the corresponding relative (normalized) time-varying frequencies by $ u_i(t) $. One has
\begin{equation}
u_i(t) \geqslant 0, \quad i \, = \, 1, 2, \ldots, n,  \label{Eq_1}
\end{equation}
\begin{equation}
\sum_{i = 1}^n u_i(t) \: = \: 1,  \label{Eq_2}
\end{equation}
i.\,e., $ \: u \, = \, (u_1, u_2, \ldots, u_n) \: $ belongs to the standard simplex in $ \mathbb{R}^n $, and
the hypercyclic dynamics is described by
\begin{equation}
\left\{ \begin{aligned}
& \frac{\mathrm{d} u_i(t)}{\mathrm{d} t} \:\: = \:\: u_i(t) \, (k_i \, u_{i - 1}(t) \: - \: f(t)), \quad
i \, = \, 1, 2, \ldots, n, \\
& f(t) \: = \: \sum_{i = 1}^n k_i \, u_i(t) \, u_{i - 1}(t), \quad u_0(t) \: = \: u_n(t), \\
& t \geqslant 0, \\
& u_i(0) \: = \: u_i^0, \quad i \, = \, 1, 2, \ldots, n.
\end{aligned} \right.  \label{Eq_3}
\end{equation}
Here $ k_i \, u_{i - 1}(t) $ is the fitness of the $ i $-th species, and $ f(t) $ is the mean fitness of the entire system
(the representation for $ f(t) $ can be obtained by adding up all of the $ n $ dynamical equations and using
the relation~(\ref{Eq_2})). Note also that the standard simplex is invariant with respect to this dynamical system.

The following properties hold:
\begin{itemize}
\item  the hypercyclic system is permanent \cite[\S 13]{HofbauerSigmund1988}, i.\,e., there exists $ \delta > 0 $
(independent from the initial coordinates $ \, u_1^0, u_2^0, \ldots, u_n^0 $) such that
$ \: u_i^0 > 0 $, $ \, i \, = \, 1, 2, \ldots, n, \: $ imply
$$
\liminf\limits_{t \, \to \, +\infty} \, u_i(t) \: > \: \delta, \quad i \, = \, 1, 2, \ldots, n;
$$
\item  for $ n = 2,3,4 $, the hypercycle admits a globally stable steady state with positive components
\cite[\S 12]{HofbauerSigmund1988}, while there appears a stable limit cycle if $ n \geqslant 5 $ \cite{MalletParetSmith1990};
\item  when $ m $ disjoint hypercycles compete with each other in the same environment, a unique hypercycle is finally
established with the other $ m - 1 $ going to extinction for almost all initial conditions \cite{Hofbauer2002};
\item  evolutionary adaptation can help hypercycles be resistant to parasites but also allows for a phase transition
phenomenon similar to the error threshold in the quasispecies models (which divides the selective phase of evolution with a clear
dominance from the random phase with a markedly more uniform distribution), as was numerically investigated in
\cite{BratusDrozhzhinYakushkina2018}.
\end{itemize}

Moreover, the paper~\cite{BratusPosvyanskii2006} studied an extension of the hypercycle model with each species distributed on
a line segment and influenced by a homogeneous diffusion, which led to a system of $ n $ partial differential equations instead of
the ordinary differential equations~(\ref{Eq_3}). This modeling approach was further developed in
\cite{BratusPosvyanskiiNovozhilov2010,BratusLukasheva2009,BratusPosvyanskiiNovozhilov2017}.

However, the actual number of macromolecules in a hypercycle may be huge, and this may significantly complicate the numerical
analysis of the associated dynamical system~(\ref{Eq_3}). It may therefore be reasonable to represent the macromolecules as
points in some line segment (of cardinality continuum) and to construct an appropriate distributed model of hypercyclic
replication. Such a methodology was previously implemented for Crow--Kimura and Eigen quasispecies models, with a single
integro-differential equation replacing a large number of ordinary differential equations
\cite{CrowKimura1964,Burger2000,BratusYegorovNovozhilov2020}. A crucial step in the construction of a hypercycle model with
a continuum of species is to incorporate the catalyzing effects (along a continuous loop in contrast to the finite closed chain
in Fig.~\ref{Fig_1}). Possible diffusive behavior is also worth taking into account. Besides, the cyclic structure has to be
ensured by stating appropriate boundary conditions at the endpoints of the line segment describing the species. For these purposes,
the current work formulates a new distributed hypercycle model based on a second-order partial integro-differential equation with
spatial delay and mixed boundary conditions. It should be emphasized that the general theory of mixed functional differential
equations including in particular partial or integro-differential equations with spatial delays is still at an early stage of
development. An introduction to this promising area of mathematical research can be found in \cite{Myshkis2005}, and some related
applications are presented in \cite{Muravnik2016,Andrianov2014,ChenMa2018}.

This paper is organized as follows. The problem is stated in Section~2. The existence, uniqueness, and nonnegativity of
the solution are discussed in Section~3, with the proofs moved to Appendix. Section~4 provides steady-state analysis.
Section~5 presents numerical simulation results for the dynamic model. Finally, concluding remarks are given in Section~6.

\section{Problem statement}

Let macromolecules (species) in a hypercycle be represented as points of the interval $ 0 \leqslant x < 2 \pi $,
and denote the relative frequency of macromolecule~$ x $ at time~$ t $ by $ u(x, t) $. The normalization condition~(\ref{Eq_2})
from the classical model transforms to
\begin{equation}
\int_0^{2 \pi} u(x, t) \, \mathrm{d} x \: = \: 1 \quad \forall t \geqslant 0,  \label{Eq_4}
\end{equation}
and it is also convenient to incorporate periodicity with respect to $ x $:
\begin{equation}
u(x + 2 \pi, \, t) \: = \: u(x, t) \quad \forall x \in \mathbb{R} \quad \forall t \geqslant 0.  \label{Eq_5}
\end{equation}
Hence, species~$ x $ and $ x + 2 n \pi $ are considered to be equivalent for any integer~$ n $. Furthermore, assume
the existence of a constant parameter~$ h \in (0, 2 \pi) $ such that replication of species~$ x $ is catalyzed by
species~$ x - h $. Next, let $ k(x) $ and $ \varphi(x) $ be functions that describe the replication rates and initial
distribution, respectively, and introduce a diffusion coefficient~$ \alpha > 0 $.

\begin{assumption}  \label{Assum_1}
$ h = \mathrm{const} \in (0, 2 \pi) $, $ \: \alpha = \mathrm{const} > 0 $, $ \: k \colon \, [0, 2 \pi] \to (0, +\infty) \: $
is a twice continuously differentiable positive function satisfying $ \, k(0) = k(2 \pi) $, $ \, k'(0) = k'(2 \pi), \, $ and
$ \: \varphi \colon \, [-h, 2 \pi] \to [0, +\infty) \: $ is a twice continuously differentiable nonnegative function
satisfying
$$
\int_0^{2 \pi} \varphi(x) \, \mathrm{d} x \: = \: 1, \quad \:
\varphi(x) \, = \, \varphi(x + 2 \pi) \quad \forall x \in [-h, 0].
$$
\end{assumption}

A hypercycle model can then be stated in the form of the following initial-boundary value problem for a second-order
partial integro-differential equation with a nonlinear source term involving a spatial delay:
\begin{equation}
\left\{ \begin{aligned}
& \frac{\partial u(x, t)}{\partial t} \:\: = \:\: u(x, t) \: (k(x) \, u(x - h, \, t) \: - \: f[u(\cdot, t)]) \:\, + \:\,
\alpha \, \frac{\partial^2 u(x, t)}{\partial x^2} \, , \\
& u(x - h, \, t) \: = \: u(x - h + 2 \pi, \, t) \quad \mathrm{if} \quad -h \leqslant x - h < 0, \\
& u(x, 0) \, = \, \varphi(x) \quad \mbox{(initial condition)}, \\
& u(0, t) \, = \, u(2 \pi, t), \quad
\frac{\partial u}{\partial x} \, (0, t) \, = \, \frac{\partial u}{\partial x} \, (2 \pi, t) \quad
\mbox{(boundary conditions)}, \\
& 0 \leqslant x \leqslant 2 \pi, \quad t \geqslant 0.
\end{aligned} \right.  \label{Eq_6}
\end{equation}
Here
\begin{equation}
f[u(\cdot, t)] \:\, = \:\, \int_0^{2 \pi} k(x) \, u(x, t) \, u(x - h, \, t) \, \mathrm{d} x \quad \forall t \geqslant 0
\label{Eq_7}
\end{equation}
is the mean fitness of the system. This representation can be obtained by integrating the dynamic equation in (\ref{Eq_6})
over $ 0 \leqslant x \leqslant 2 \pi $ and using the normalization property~(\ref{Eq_4}) as well as the boundary condition
for $ \partial u / \partial x $. Similar arguments can also help us verify that, under Assumption~\ref{Assum_1}, a solution of
(\ref{Eq_6}) should satisfy
$$
\frac{\mathrm{d}}{\mathrm{d} t} \left( \int_0^{2 \pi} u(x, t) \, \mathrm{d} x \right) \:\, = \:\, 0 \quad
\forall t \geqslant 0,
$$
which ensures (\ref{Eq_4}). Moreover, the mixed boundary conditions come from the aforementioned periodicity (cyclic
structure) with respect to $ x $. Thus, the relations~(\ref{Eq_6}) and (\ref{Eq_7}) can serve as a distributed modification
of the classical hypercycle model~(\ref{Eq_3}).

Note that the catalytic structure leading to a spatial delay is an important feature of our distributed hypercycle model
in comparison with distributed quasispecies models \cite{CrowKimura1964,Burger2000,BratusYegorovNovozhilov2020}.

\section{Existence, uniqueness, and nonnegativity of the solution}

This section addresses the existence, uniqueness, and nonnegativity of the classical solution to the initial-boundary value
problem~(\ref{Eq_6}), with detailed proofs moved to Appendix.

Introduce an arbitrary time horizon~$ T > 0 $ and the Banakh space $ \, \mathbf{C}^{2, 1} ([0, 2 \pi] \times [0, T]) \, $ consisting of
continuous functions $ \: [0, 2 \pi] \times [0, T] \, \ni \, (x, t) \: \longmapsto \: w(x, t) \, \in \, \mathbb{R} \: $
such that $ \: \partial w / \partial x $, $ \partial^2 w / \partial x^2 $, $ \partial w / \partial t \: $ are continuous on
$ \, [0, 2 \pi] \times [0, T] $. The continuity on the boundary of the rectangle means that the corresponding bounded limits
from the interior exist and are taken as the values on the boundary. The norm in
$ \, \mathbf{C}^{2, 1} ([0, 2 \pi] \times [0, T]) \, $ is defined as
\begin{equation}
\begin{aligned}
\| w \|_{\mathbf{C}^{2, 1} ([0, 2 \pi] \times [0, T])} \:\: = \:\: \max_{ \substack{0 \leqslant x \leqslant 2 \pi, \\
0 \leqslant t \leqslant T} } \: \left\{ |w(x, t)| \: + \: \left| \frac{\partial w(x, t)}{\partial x} \right| \right. & \\
\left. + \: \left| \frac{\partial^2 w(x, t)}{\partial x^2} \right| \: + \:
\left| \frac{\partial w(x, t)}{\partial t} \right| \right\}.
\end{aligned}  \label{Eq_8}
\end{equation}
The classical solution is searched for in this space.

The space~$ \mathbf{C}^2 ([0, 2 \pi]) $ is similarly defined (but with excluded dependence on~$ t $).

Consider the auxiliary linear problem
\begin{equation}
\left\{ \begin{aligned}
& \mathcal{A} v(x, t) \:\: = \:\: \frac{\partial v(x, t)}{\partial t} \: - \: \alpha \,
\frac{\partial^2 v(x, t)}{\partial x^2} \:\: = \:\: \psi(x, t), \\
& v(x, 0) \, = \, \varphi(x) \quad \mbox{(initial condition)}, \\
& v(0, t) \, = \, v(2 \pi, t), \quad
\frac{\partial v}{\partial x} \, (0, t) \, = \, \frac{\partial v}{\partial x} \, (2 \pi, t) \quad
\mbox{(boundary conditions)}, \\
& 0 \leqslant x \leqslant 2 \pi, \quad 0 \leqslant t \leqslant T,
\end{aligned} \right.  \label{Eq_9}
\end{equation}
with a right-hand side~$ \psi $ that lies in $ \, \mathbf{C}^{2, 1} ([0, 2 \pi] \times [0, T]) \, $ and satisfies
the boundary conditions
\begin{equation}
\psi(0, t) \, = \, \psi(2 \pi, t), \quad
\frac{\partial \psi}{\partial x} \, (0, t) \, = \, \frac{\partial \psi}{\partial x} \, (2 \pi, t) \quad \:
\forall t \in [0, T].  \label{Eq_10}
\end{equation}
The corresponding solution is obtained in Appendix~\ref{App_1}, and, with the help of Assumption~\ref{Assum_1},
one can directly verify that it also belongs to $ \, \mathbf{C}^{2, 1} ([0, 2 \pi] \times [0, T]) $.
The relations~(\ref{Eq_A_1}), (\ref{Eq_A_4}), (\ref{Eq_A_5}) give rise to
the operators~$ \mathcal{E}, \mathcal{E}_1, \mathcal{E}_2 $ such that
\begin{equation}
v(x, t) \: = \: \mathcal{E} [\varphi, \psi] (x, t) \: = \: \mathcal{E}_1 \varphi(x, t) \, + \, \mathcal{E}_2 \psi(x, t).
\label{Eq_11}
\end{equation}
Note that $ \mathcal{E}_1, \mathcal{E}_2 $ are linear operators. Straightforward calculations based on (\ref{Eq_A_1}),
(\ref{Eq_A_4}), (\ref{Eq_A_5}) lead to the estimates
\begin{equation}
\begin{aligned}
& \| \mathcal{E}_1 \varphi \|_{\mathbf{C}^{2, 1} ([0, 2 \pi] \times [0, T])} \:\: \leqslant \:\:
c_1 \: \| \varphi \|_{\mathbf{C}^2 ([0, 2 \pi])} \, , \\
& \| \mathcal{E}_2 \psi \|_{\mathbf{C}^{2, 1} ([0, 2 \pi] \times [0, T])} \:\: \leqslant \:\:
c_2 \: \| \psi \|_{\mathbf{C}^{2, 1} ([0, 2 \pi] \times [0, T])} \, , \\
& \| \mathcal{E} [\varphi, \psi] \|_{\mathbf{C}^{2, 1} ([0, 2 \pi] \times [0, T])} \:\: \leqslant \:\:
c_1 \: \| \varphi \|_{\mathbf{C}^2 ([0, 2 \pi])} \:\, + \:\, c_2 \:
\| \psi \|_{\mathbf{C}^{2, 1} ([0, 2 \pi] \times [0, T])}
\end{aligned}  \label{Eq_12}
\end{equation}
with some positive constants~$ c_1, c_2 $.

The idea for the proof of an existence and uniqueness result for (\ref{Eq_6}) is to apply the Banach fixed point theorem
(see, e.\,g., \cite[Theorem~4.16]{Muscat2014}) to the iterative process
\begin{equation}
u^0(x, t) \: = \: \varphi(x), \quad u^{m + 1} \: = \: \mathcal{E} [\varphi, \, \mathcal{F} u^m], \quad m \, = \, 0, 1, 2, \ldots,
\label{Eq_13}
\end{equation}
where
\begin{equation}
\begin{aligned}
& \mathcal{F} u^m(x, t) \:\, = \:\, u^m(x, t) \: (k(x) \, u^m(x - h, \, t) \: - \: f[u^m(\cdot, t)]), \\
& f[u^m(\cdot, t)] \:\, = \:\, \int_0^{2 \pi} k(x) \, u^m(x, t) \, u^m(x - h, \, t) \, \mathrm{d} x
\end{aligned}  \label{Eq_14}
\end{equation}
(see the right-hand side of the dynamic equation in (\ref{Eq_6}), as well as the mean fitness definition~(\ref{Eq_7})).
Because of the nonlinearity in (\ref{Eq_14}), this proof approach works only under the quite restrictive technical condition
that the quantity
\begin{equation}
\bar{k} \:\: = \:\: \max_{0 \leqslant x \leqslant 2 \pi} \: \max \, \{ |k(x)|, \, |k'(x)|, \, |k''(x)| \}  \label{Eq_15}
\end{equation}
is sufficiently small. The proof of an extended existence and uniqueness result remains an open task.
One may in general expect the unique solution to exist under weaker conditions.

Due to the boundary conditions for $ \varphi, \varphi', k, k' $ at $ x = 0 $ and $ x = 2 \pi $ (recall
Assumption~\ref{Assum_1}), the relations~(\ref{Eq_10}) should hold with $ \psi = \mathcal{F} u^m $ for all $ m = 0, 1, 2, \ldots $
Using also the fact that $ \: \int_0^{2 \pi} u^0(x, t) \, \mathrm{d} x \: = \: \int_0^{2 \pi} \varphi(x) \, \mathrm{d} x \: = \: 1, \: $
one arrives at
\begin{equation}
\int_0^{2 \pi} u^m(x, t) \, \mathrm{d} x \:\, = \:\, \int_0^{2 \pi} \mathcal{F} u^m(x, t) \, \mathrm{d} x \:\, = \:\, 1, \quad
m \, = \, 0, 1, 2, \ldots  \label{Eq_16}
\end{equation}

The following lemma is established in Appendix~\ref{App_2}.

\begin{lemma}  \label{Lemma_2}
Under Assumption~\ref{Assum_1}, the sequence of the norms $ \: \| u^m \|_{\mathbf{C}^{2, 1} ([0, 2 \pi] \times [0, T])} $,
$ \, m = 0, 1, 2, \ldots, \: $ is uniformly bounded for sufficiently small values of (\ref{Eq_15}).
\end{lemma}

The proof of the existence and uniqueness result (Theorem~\ref{Thm_3}) is given in Appendix~\ref{App_3}.

\begin{theorem}  \label{Thm_3}
Let Assumption~\ref{Assum_1} hold. Then for for sufficiently small values of (\ref{Eq_15}), there exists a unique classical
solution $ \: u \, \in \, \mathbf{C}^{2, 1} ([0, 2 \pi] \times [0, T]) \: $ to the distributed hypercycle problem~(\ref{Eq_6}).
\end{theorem}

The solution nonnegativity result (Theorem~\ref{Thm_4}) is proved in Appendix~\ref{App_4} under the condition that
the catalytic shift (delay) parameter~$ h $ is sufficiently small. This condition seems reasonable when comparing
the distributed hypercycle model~(\ref{Eq_6}),(\ref{Eq_7}) to the classical model~(\ref{Eq_3}) with a large number of
species~$ n $. Indeed, the relation between $ h $ and $ n $ can be set up to $ \, h = 2 \pi / n, \, $ so that $ h \to 0 $ as
$ n \to \infty $.

\begin{theorem}  \label{Thm_4}
Let Assumption~\ref{Assum_1} hold, and let a solution $ \: u_h \, \in \, \mathbf{C}^{2, 1} ([0, 2 \pi] \times [0, T]) \: $
to the problem~(\ref{Eq_6}) exist for all $ \, 0 < h \leqslant h_0, \, $ where $ \: h_0 = \mathrm{const} \in (0, 2 \pi) $. Assume also
\begin{equation}
\sup_{0 < h \leqslant h_0} \, \| u_h \|_{\mathbf{C}^{2, 1} ([0, 2 \pi] \times [0, T])} \:\, < \:\, \infty.  \label{Eq_17}
\end{equation}
Then $ u_h $ is a nonnegative function for sufficiently small~$ h $.
\end{theorem}

\begin{remark}  \label{Rem_5}  \rm
Let Assumption~\ref{Assum_1} hold, and let $ \bar{k} $ be small enough to apply Lemma~\ref{Lemma_2} and Theorem~\ref{Thm_3}.
First, note that $ c_1, c_2 $ in the estimates~(\ref{Eq_12}) do not depend on $ h $, since the auxiliary problem~(\ref{Eq_9})
does not contain $ h $. From the proofs of Lemma~\ref{Lemma_2} and Theorem~\ref{Thm_3} in Appendix, one can see that
the parameters in the corresponding estimates also do not depend on $ h $. Hence, the condition~(\ref{Eq_17}) is fulfilled in
this case.
\end{remark}

\section{Steady-state analysis}
System \eqref{Eq_6}, \eqref{Eq_7} has the following equilibrium solution $v(x)=(2\pi)^{-1}$ for $k(x)=k=\text{const}>0$. First we show that this equilibrium is unstable.

Let $\psi\in\mathbf C^{2,1}$. Consider function
\begin{equation}\label{Eq_18}
    v_\varepsilon(x,t)=\frac{1}{2\pi}+\varepsilon \psi(x,t)
\end{equation}
that satisfies condition \eqref{Eq_4}. Then
\begin{equation}\label{Eq_19}
    \int_0^{2\pi}\psi(x,t)\mathrm{d}x=0.
\end{equation}
Plugging $v_\varepsilon$ into \eqref{Eq_6} and keeping only the linear with respect to $\varepsilon$ terms we obtain
\begin{align*}
\frac{\partial \psi}{\partial t}(x,t)= & k\bigl(\psi(x,t)+\psi(x-h,t)\bigr)-k\psi(x,t)\\
&-\frac{k}{2\pi}\int_0^{2\pi}\bigl(\psi(x,t)-\psi(x-h,t)\bigr)\mathrm{d}x+\alpha\frac{\partial^2\psi}{\partial x^2}(x,t),\\
\psi(x,0)= & \psi_0(x),\quad x\in[0,2\pi],
\end{align*}
which, after simplification and taking into account \eqref{Eq_19}, leads to
\begin{equation}\label{Eq_20}
\begin{split}
  \frac{\partial \psi}{\partial t}(x,t)=&k \psi(x-h,t)+\alpha \frac{\partial^2\psi}{\partial x^2}(x,t),\quad t>0, \\
  \psi(x,0)=&\psi_0(x), x\in[0,2\pi].
\end{split}
\end{equation}
It is natural to look for a solution to this problem in the form of a Fourier series
$$
\psi(x,t)=\sum_{n=-\infty}^{+\infty}c_n(t)e^{-inx},
$$
which yields
\begin{equation}\label{Eq_21}
    \frac{\mathrm{d}c_n}{\mathrm{d}t}(t)=c_n(t)\left(ke^{inh}-\alpha n^2\right),\quad n=0,\pm 1,\pm 2,\ldots.
\end{equation}
For $n=0$ $c_0(t)\to\infty$ for $t\to\infty$ and hence we proved that the equilibrium $v(x)=(2\pi)^{-1}$ is unstable.

Now we look into possible spatially non-homogeneous equilibria, which must satisfy the following problem
\begin{equation}\label{Eq_22}
\begin{split}
  \alpha \frac{\mathrm{d}^2 v}{\mathrm{d} x^2}(x) & + v(x)\bigl(k v(x-h)-\bar{f})\bigr) =0,\quad x\in(0,2\pi), \\
   v'(0) &=v'(2\pi),\quad \bar{f}=k\int_0^{2\pi}v(x)v(x-h)\mathrm{d}x,\\
   v(x-h)&=\begin{cases}
   v(x-h),&\text{if }x-h\geqslant 0,\\
   v(2\pi+x-h),&\text{if } x-h<0,
   \end{cases}\\
   \int_0^{2\pi}v(x)\mathrm{d}x&=\int_0^{2\pi}v(x-h)\mathrm{d}x=1.
\end{split}
\end{equation}
As before, here we assume that $k(x)=k=\text{const}>0$.

If we assume that $v\in\mathbf C^2[0,2\pi]$ then
$$
v(x-h)=v(x)-hv_{x}(x)+\frac{h^2}{2}v_{xx}+o(h^2).
$$
Using this representation and keeping only the terms up to $o(h^2)$ the first equation in \eqref{Eq_22} becomes
$$
v_{xx}(x)\left(\alpha+\frac{kh^2}{2}v(x)\right)=\bar{f}v(x)-kv(x)\bigl(v(x)-hv_x(x)\bigr).
$$
We rewrite this second order equation as the following system:
\begin{equation}\label{Eq_23}
\begin{split}
    \frac{\mathrm{d} v}{\mathrm d x}(x)&=w(x),\\
    \frac{\mathrm{d} w}{\mathrm d x}(x)&=\frac{\bigl(\bar f-kv(x)+khw(x)\bigr) v(x)}{\alpha+2^{-1}kh^2v(x)}\,,
\end{split}
\end{equation}
with the boundary conditions $w(0)=w(2\pi)$.

The standard analysis of \eqref{Eq_23} yields that there are two equilibria $O(0,0)$ and $A(\bar fk^{-1},0)$. The former one is a saddle point for any parameter values, and the latter one is a stable focus for $h<0$, unstable focus for $h>0$; if $h=0$ then the eigenvalues of the Jacobi matrix are pure imaginary complex conjugate.

\begin{figure}
\includegraphics[width=0.45\textwidth]{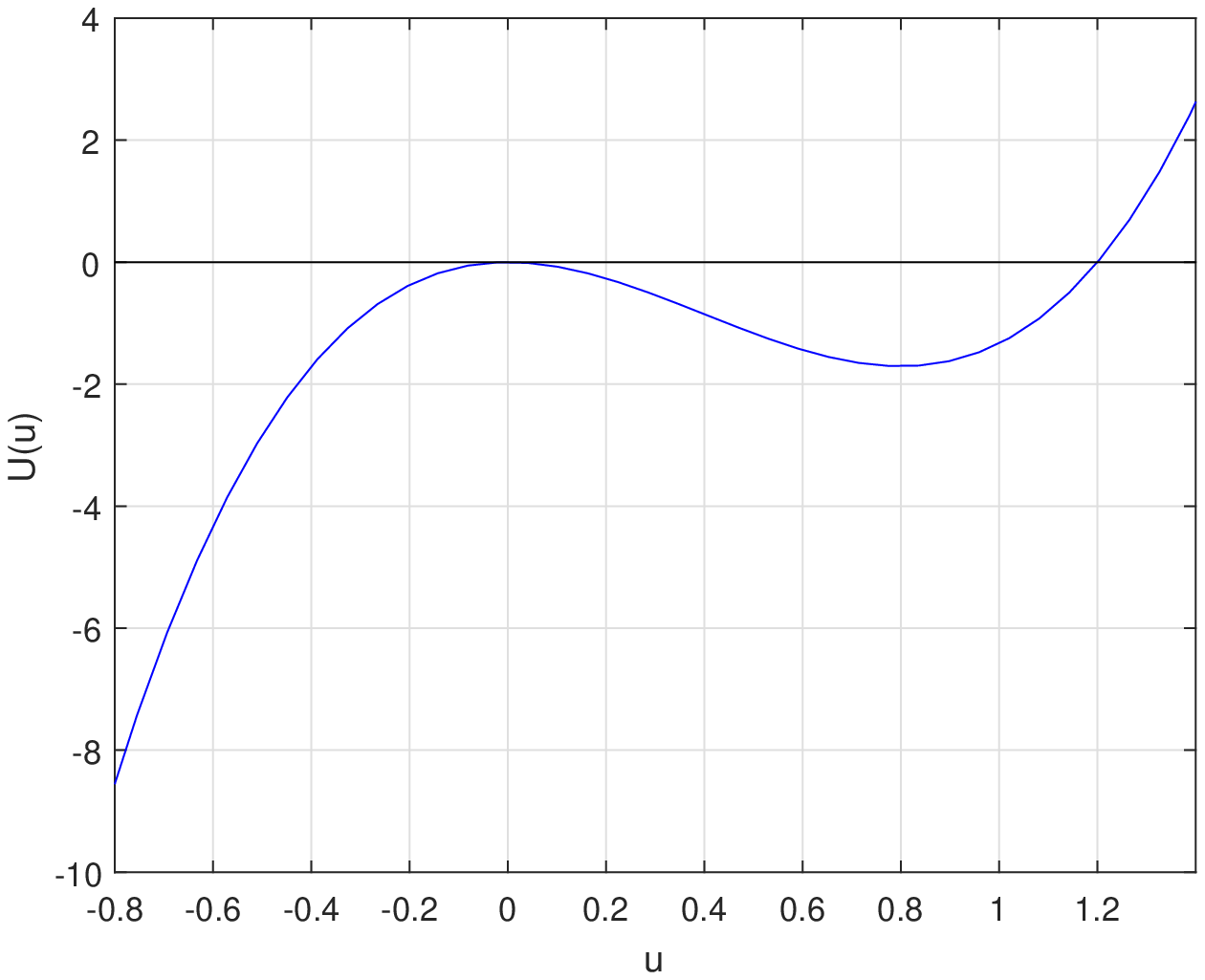}
\includegraphics[width=0.54\textwidth]{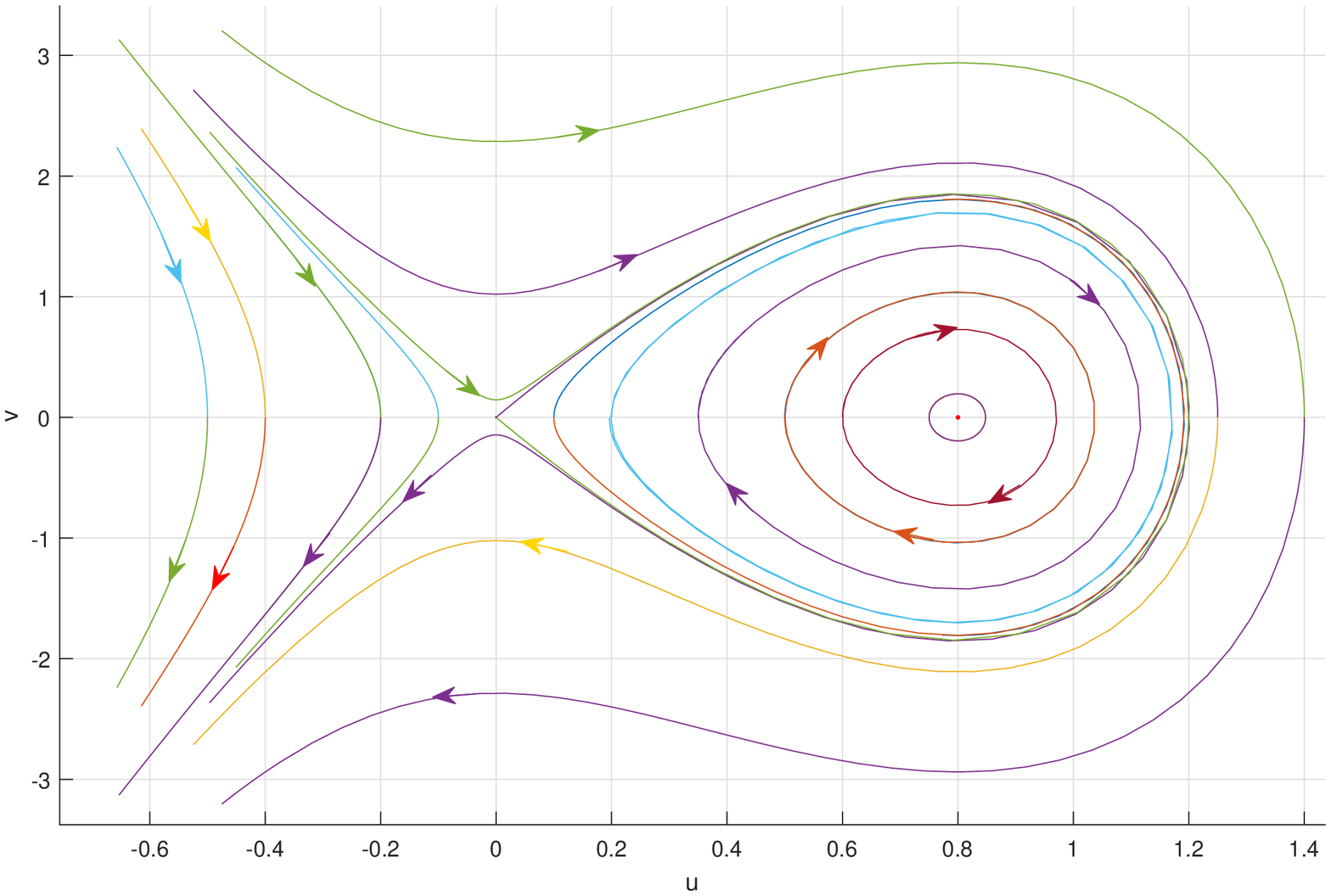}
\caption{Graph of the potential for system \eqref{Eq_23} with $h=0$ (left) and the corresponding phase portrait (right).}\label{Fig_2}
\end{figure}

We note that for $h=0$ system \eqref{Eq_23} becomes Hamiltonian, with the Hamiltonian
$$
H(v,w)=\frac{w^2}{2}+U(v),\quad U(v)=\frac{v^2}{\alpha}\left(\frac{kv}{2}-\frac{\bar f}{2}\right).
$$
The graph of potential $U$ immediately implies that equilibrium $A$ in the case $h=0$ is Lyapunov stable, with a family of closed orbits surrounding it, which implies that there are infinitely many orbits that satisfy the  condition $w(x_1)=w(x_2)=0$ for two points $x_2>x_1>0$ (see Fig. \ref{Fig_2}).

\begin{figure}
\centering
\includegraphics[width=0.65\textwidth]{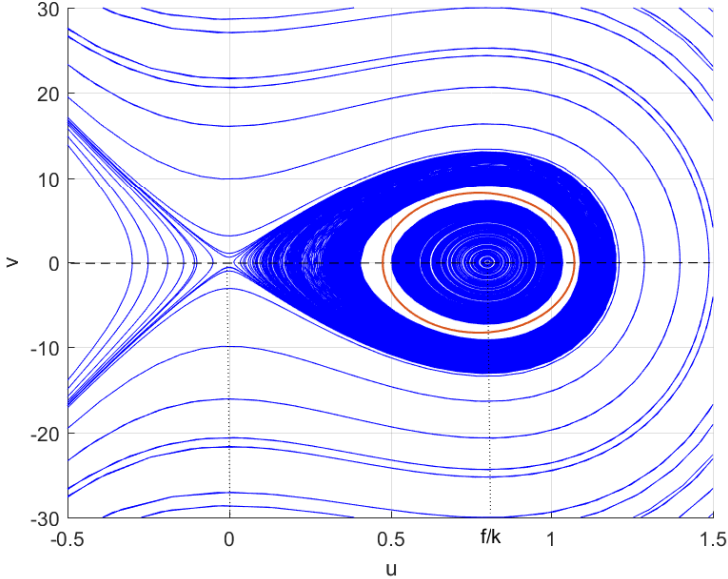}
\caption{Periodic solution of \eqref{Eq_23} for sufficiently small $\alpha$ and $h>0$.}\label{Fig_3}
\end{figure}

The conditions identified above are sufficient to invoke the Hopf bifurcation theorem (in the form given, e.g., in \cite{marsden2012hopf}, Section 3C) to conclude that for sufficiently small $|h|$ there must be nonconstant periodic solutions that collapse at a point when $|h|\to 0$. Among all such periodic solutions we are interested in one that satisfies $w(0)=w(2\pi)=0$. We claim that such solution can always be found if one is allowed to consider sufficiently small values of $\alpha$ (see Fig. \ref{Fig_3}).

Indeed, from equations \eqref{Eq_23} we see that decreasing $\alpha=o(h)$ leads to increase of velocity of movements along the phase portrait, and therefore there will always be sufficiently small values of $\alpha$ for which the required condition is satisfied; moreover, it is clear that for some choice of $\alpha$ the orbit will travel around the equilibrium point only once, for some $\alpha$ will travel twice, etc. As a result of these reasoning we put forward the following conjecture.
\begin{conjecture}
For sufficiently small values of $h$ and diffusion coefficient $\alpha=o(h)$ there exist periodic solutions to \eqref{Eq_22}, with are $o(h^2)$-close to the periodic solutions to \eqref{Eq_23}.
\end{conjecture}

An additional support to this conjecture can be seen from the following argument.

If we assume that $v$ is sufficiently smooth in \eqref{Eq_22}, i.e.,
$$
v(x-h)=\sum_{j=0}^N\frac{(-1)^j}{j!}h^jv^{(j)}(x)+o(h^N),
$$
then problem \eqref{Eq_22} can be rewritten as a system of $N$ (dropping the terms of the smaller order) equations of the first order
\begin{equation}\label{Eq_24}
\begin{split}
\frac{\mathrm{d} v_1}{\mathrm{d} x}(x)&=v_2(x),\quad \frac{\mathrm{d} v_2}{\mathrm{d} x}(x)=v_3(x),\quad \ldots,\quad \frac{\mathrm{d} v_{N-1}}{\mathrm{d} x}(x)=v_{N}(x),\\
\frac{\mathrm{d} v_N}{\mathrm{d} x}(x)&=\frac{(-1)^{N+1}N!}{kh^N}\left(k\sum_{j=0}^{N-1}\frac{(-1)^j}{j!}h^j v_{j+1}+\alpha\frac{v_3}{v_1}-\bar f\right),
\end{split}
\end{equation}
which has the equilibrium $A(\bar fk^{-1},0,\ldots,0)$. Linearizing around this equilibrium yield the Jacobi matrix with the characteristic polynomial
$$
(-1)^{N+1}\frac{kh^N}{N!}\lambda^N+(-1)^{N-1}\frac{h^{N-1}}{(N-1)!}\lambda^{N-1}+\ldots-\frac{h^3}{3!}\lambda^3+\left(\frac{h^2}{2!}+\frac{\alpha}{\bar f}\right)\lambda^2-h\lambda+1.
$$
When $h\to 0$ we again have two purely imaginary complex root, and the same Hopf theorem \cite{marsden2012hopf} implies the existence of periodic solutions that collapse into the point when $h\to 0$.

Therefore, it is highly probably that problem \eqref{Eq_22} admits solutions different from a constant, which may be non-unique; and therefore the original system \eqref{Eq_6}, \eqref{Eq_7} admits spatially non-uniform equilibria.
\section{Numerical solutions}\label{sec_5}
In the previous section we found a family of spatially-nonhomogeneous stationary solution. We did not study their stability analytically because it represents an independent and complex problem on its own. Here we present some numerical evidence that these solutions are most probably unstable because in numerical experiments spatial and temporal oscillations are usually observed.

\begin{figure}
\centering
\includegraphics[width=0.7\textwidth]{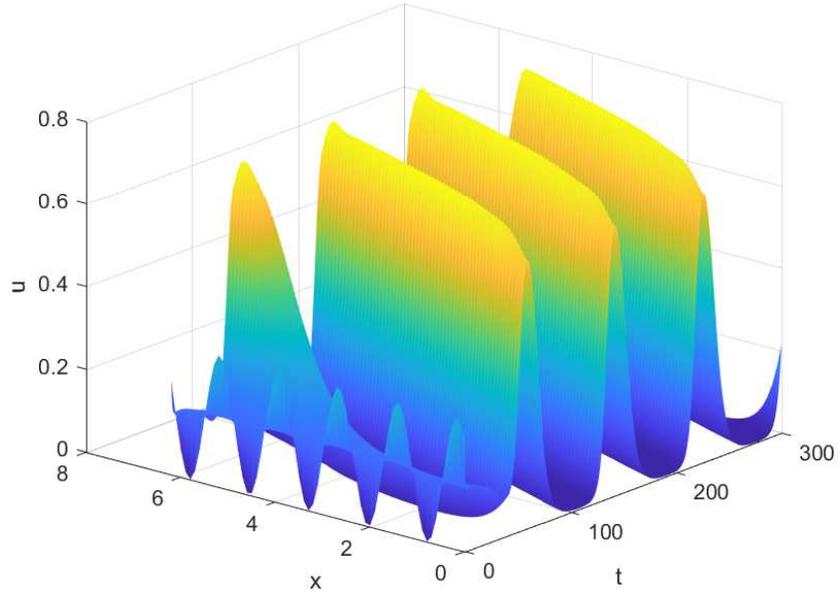}
\caption{Numerical solution of problem \eqref{Eq_6} with $x\in[0,2\pi],\,\alpha=0.05,t\in[0,300],h=0.4, k(x)=1,\varphi(x)=(2\pi)^{-1}(\sin(5x+\pi/4)+1)$.}\label{fig_4}
\end{figure}

To conduct the numerical experiments we use an explicit finite difference scheme and numerical approximation of the integral on the given interval. Choosing the parameters as follows: $x\in[0,2\pi],\,\alpha=0.05,t\in[0,300],h=0.4, k(x)=1,\varphi(x)=(2\pi)^{-1}(\sin(5x+\pi/4)+1)$ we can see that the numerically obtained solution $u$ represents a non-linear wave which oscillates with respect to both spatial and temporal variables (see Fig. \ref{fig_4}).

\begin{figure}
\centering
\includegraphics[width=0.7\textwidth]{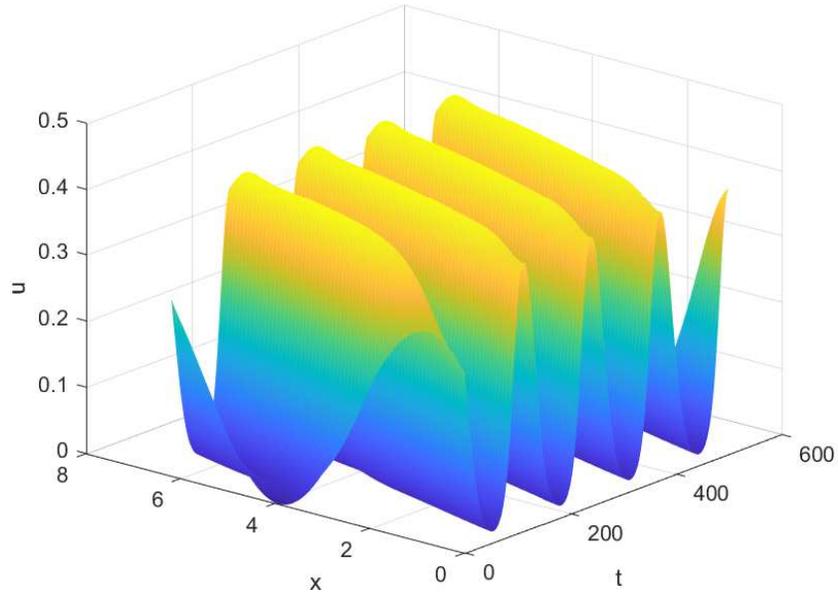}
\caption{Numerical solution of problem \eqref{Eq_6} with $x\in[0,2\pi],\,\alpha=0.05,t\in[0,500],h=0.4, k(x)=1,\varphi(x)=(2\pi)^{-1}(\sin(x+\pi/4)+1)$.}\label{fig_5}
\end{figure}

Choosing a different initial condition does not change the result qualitatively (see Fig. \ref{fig_5}).
Among other things we can see that the numerical simulations strongly indicate that the spatially continuous hypercyclic system \eqref{Eq_6} is permanent similar to its discrete counterpart. It is an interesting open mathematical problem to analyze the observed oscillations analytically.

\section{Conclusion}In this paper we introduced a continuous analogue of the classical discrete hypercycle \eqref{Eq_3}. The dynamics of \eqref{Eq_3} is well understood; it is, however, an open challenge to fully investigate the properties of the solutions of the introduced partial integro-differential equation with spatial delay \eqref{Eq_6}. Here we started such investigation by proving the existence and uniqueness theorem for this problem and showing, according to biological interpretation of the model, non-negativity of its solutions. We also presented some analytical evidence that there are spatially nonuniform stationary solutions and numerical evidence that the actual asymptotic behavior of the system tends to both spatial and temporal oscillations. It is a challenging open problem to provide an analytical support for our observations.

We would like to conclude this paper with a short historic anecdote. Famous Soviet mathematician Anatolii Myshkis (1920--2009) is mostly known as one of the founders of the field of delay differential equations. Less known is the fact that the last significant mathematical object of his study was a class of differential equations with delay with respect to other than time variable \cite{Myshkis2005}. Two of the authors of the present paper were fortunate to attend a talk by prof. Myshkis in early 2000's devoted to this topic. As it was usual in mathematical presentations in Soviet Union and later in Russia, the talk concluded with a somewhat long discussion of the results. In particular prof. Myshkis mentioned that the class of problem he had been studying did not get much attention in the mathematical community in part because of the lack of applications and mathematical models of natural phenomena described by such equations. Several ideas were suggested in which direction one may look for applications, but none of them was very convincing. For us it is important to state that the mathematical model \eqref{Eq_6} we consider in this text is very similar to the mixed functional differential equations considered in \cite{Myshkis2005} and it is our hope that our model will attract more attention to this interesting class of mathematical problems.
\section*{Acknowledgements}

The work of ASB was supported by grant $075-15-219-1621$ from Ministry of science and higher education of the Russian Federation.



\appendix

\section*{Appendix}

\setcounter{equation}{0}
\renewcommand{\theequation}{A.\arabic{equation}}
\renewcommand{\thesubsection}{A.\arabic{subsection}}

\subsection{Solution of the auxiliary linear problem~(\ref{Eq_9})}  \label{App_1}

The boundary conditions are obviously satisfied if one represents the solution as the Fourier series
\begin{equation}
v(x, t) \:\: = \:\: \frac{a_0(t)}{2} \: + \: \sum_{n = 1}^{\infty} (a_n(t) \cos(n x) \, + \, b_n(t) \sin(n x)).
\label{Eq_A_1}
\end{equation}
Similarly, the initial profile and the right-hand side of the equation are written as
\begin{equation}
\varphi(x) \:\: = \:\: \frac{\varphi_{10}}{2} \: + \: \sum_{n = 1}^{\infty} (\varphi_{1n} \cos(n x) \, + \,
\varphi_{2n} \sin(n x)),
\label{Eq_A_2}
\end{equation}
\begin{equation}
\psi(x, t) \:\: = \:\: \frac{\psi_{10}(t)}{2} \: + \: \sum_{n = 1}^{\infty} (\psi_{1n}(t) \cos(n x) \, + \,
\psi_{2n}(t) \sin(n x)).
\label{Eq_A_3}
\end{equation}
Here we use the fact that the functional system $ \: 1 / \sqrt{2 \pi} $, $ \: \left( 1 / \sqrt{\pi} \right) \cos(n x) $,
$ \: \left( 1 / \sqrt{\pi} \right) \sin(n x) $, $ \: n = 1, 2, 3, \ldots, \: $ is orthonormal in the Hilbert
space~$ \, \mathcal{L}^2 ([0, 2 \pi]; \, \mathbb{R}) \, $ of square Lebesgue-integrable real-valued functions on $ [0, 2 \pi] $.
The Fourier coefficients in (\ref{Eq_A_2}) and (\ref{Eq_A_3}) are determined by
$$
\begin{aligned}
& \varphi_{10} \:\, = \:\, \frac{1}{\pi} \, \int_0^{2 \pi} \varphi(x) \, \mathrm{d} x, \quad
\psi_{10}(t) \:\, = \:\, \frac{1}{\pi} \, \int_0^{2 \pi} \psi(x, t) \, \mathrm{d} x, \\
& \varphi_{1n} \:\, = \:\, \frac{1}{\pi} \, \int_0^{2 \pi} \varphi(x) \cos(n x) \, \mathrm{d} x, \\
& \varphi_{2n} \:\, = \:\, \frac{1}{\pi} \, \int_0^{2 \pi} \varphi(x) \sin(n x) \, \mathrm{d} x, \\
& \psi_{1n}(t) \:\, = \:\, \frac{1}{\pi} \, \int_0^{2 \pi} \psi(x, t) \cos(n x) \, \mathrm{d} x, \\
& \psi_{2n}(t) \:\, = \:\, \frac{1}{\pi} \, \int_0^{2 \pi} \psi(x, t) \sin(n x) \, \mathrm{d} x, \\
& n \, = \, 1, 2, 3, \ldots
\end{aligned}
$$
Using integration by parts together with the conditions $ \, \varphi(0) = \varphi(2 \pi) $, $ \, \varphi'(0) = \varphi'(2 \pi) \, $
(recall Assumption~\ref{Assum_1}), and (\ref{Eq_10}), one arrives at the following representations:
\begin{equation}
\begin{aligned}
& \varphi_{1n} \:\, = \:\, \frac{1}{n \pi} \, \int_0^{2 \pi} \varphi(x) \, \mathrm{d} (\sin(n x)) \:\, = \:\,
-\frac{1}{n \pi} \, \int_0^{2 \pi} \varphi'(x) \sin(n x) \, \mathrm{d} x \\
& \qquad
= \:\, \frac{1}{n^2 \pi} \, \int_0^{2 \pi} \varphi'(x) \, \mathrm{d} (\cos(n x)) \:\, = \:\,
-\frac{1}{n^2 \pi} \, \int_0^{2 \pi} \varphi''(x) \cos(n x) \, \mathrm{d} x, \\
& \varphi_{2n} \:\, = \:\, -\frac{1}{n \pi} \, \int_0^{2 \pi} \varphi(x) \, \mathrm{d} (\cos(n x)) \:\, = \:\,
\frac{1}{n \pi} \, \int_0^{2 \pi} \varphi'(x) \cos(n x) \, \mathrm{d} x \\
& \qquad
= \:\, \frac{1}{n^2 \pi} \, \int_0^{2 \pi} \varphi'(x) \, \mathrm{d} (\sin(n x)) \:\, = \:\,
-\frac{1}{n^2 \pi} \, \int_0^{2 \pi} \varphi''(x) \sin(n x) \, \mathrm{d} x, \\
& \psi_{1n}(t) \:\, = \:\, -\frac{1}{n^2 \pi} \, \int_0^{2 \pi} \frac{\partial^2 \psi(x, t)}{\partial x^2} \,
\cos(n x) \, \mathrm{d} x, \\
& \psi_{2n}(t) \:\, = \:\, -\frac{1}{n^2 \pi} \, \int_0^{2 \pi} \frac{\partial^2 \psi(x, t)}{\partial x^2} \,
\sin(n x) \, \mathrm{d} x, \\
& n \, = \, 1, 2, 3, \ldots
\end{aligned}  \label{Eq_A_4}
\end{equation}

Plugging the relations~(\ref{Eq_A_1}), (\ref{Eq_A_3}) into the equation of (\ref{Eq_9}) yields
$$
\dot{a}_0(t) \: = \: \psi_{10}(t), \quad \dot{a}_n(t) \, + \, \alpha n^2 a_n(t) \: = \: \psi_{1n}(t), \quad
\dot{b}_n(t) \, + \, \alpha n^2 b_n(t) \: = \: \psi_{2n}(t),
$$
while the initial condition with (\ref{Eq_A_1}), (\ref{Eq_A_2}) implies
$$
a_0(0) \: = \: \varphi_{10}, \quad a_n(0) \: = \: \varphi_{1n}, \quad b_n(0) \: = \: \varphi_{2n},
$$
and one consequently obtains the sought-after Fourier coefficients:
\begin{equation}
\begin{aligned}
& a_0(t) \:\: = \:\: \varphi_{10} \: + \: \int_0^t \psi_{10}(s) \, \mathrm{d} s,  \\
& a_n(t) \:\: = \:\: e^{-\alpha n^2 t} \, \varphi_{1n} \: + \: \int_0^t e^{-\alpha n^2 (t - s)} \, \psi_{1n}(s) \, \mathrm{d} s,  \\
& b_n(t) \:\: = \:\: e^{-\alpha n^2 t} \, \varphi_{2n} \: + \: \int_0^t e^{-\alpha n^2 (t - s)} \, \psi_{2n}(s) \, \mathrm{d} s, \\
& n \, = \, 1, 2, 3, \ldots
\end{aligned}  \label{Eq_A_5}
\end{equation}

\subsection{Proof of Lemma~\ref{Lemma_2}}  \label{App_2}

From (\ref{Eq_14})--(\ref{Eq_16}), one derives
$$
\mathcal{F} u^m(x, t) \:\, = \:\, u^m(x, t) \: \int_0^{2 \pi} (k(x) \, u^m(x - h, \, t) \: - \:
k(\xi) \, u^m(\xi - h, \, t)) \: u^m(\xi, t) \: \mathrm{d} \xi
$$
and, therefore,
$$
\| \mathcal{F} u^m \|_{\mathbf{C}^{2, 1} ([0, 2 \pi] \times [0, T])} \:\, \leqslant \:\,
c_3 \bar{k} \, \| u^m \|^3_{\mathbf{C}^{2, 1} ([0, 2 \pi] \times [0, T])} \, , \quad
m \, = \, 0, 1, 2, \ldots,
$$
where $ \, c_3 = \mathrm{const} > 0 $. Together with (\ref{Eq_12}) and (\ref{Eq_13}), this yields
\begin{equation}
\begin{aligned}
& \left\| u^{m + 1} \right\|_{\mathbf{C}^{2, 1} ([0, 2 \pi] \times [0, T])} \:\, \leqslant \:\,
c_4 \: + \: c_2 c_3 \bar{k} \, \| u^m \|^3_{\mathbf{C}^{2, 1} ([0, 2 \pi] \times [0, T])} \, , \\
& m \, = \, 0, 1, 2, \ldots, \quad c_4 \: = \: c_1 \, \| \varphi \|_{\mathbf{C}^2 ([0, 2 \pi])} \, , \quad
u^0(x, t) \: = \: \varphi(x).
\end{aligned}  \label{Eq_A_6}
\end{equation}

We need the following auxiliary algebraic properties for nonnegative parameters:
\begin{equation}
\begin{aligned}
& (p_1 + p_2)^3 \:\: = \:\: p_1^3 \: + \: 3 p_1^2 p_2 \: + \: 3 p_1 p_2^2 \: + \: p_2^3 \\
& \qquad\qquad \:\:\:\:
\leqslant \:\: p_1^3 \: + \: 6 \, (\max \, \{ p_1, p_2 \})^3 \: + \: p_2^3 \\
& \qquad\qquad \:\:\:\:
\leqslant \:\: p_1^3 \: + \: 6 \left( p_1^3 + p_2^3 \right) \: + \: p_2^3 \:\: \leqslant \:\:
7 \left( p_1^3 + p_2^3 \right), \\
& (p_1 + p_2 + p_3)^3 \:\: \leqslant \:\: 7 p_1^3 \: + \: 7 (p_2 + p_3)^3 \:\: \leqslant \:\:
7 p_1^3 \: + \: 7^2 p_2^3 \: + \: 7^2 p_3^3, \\
& (p_1 + p_2 + \ldots + p_l)^3 \:\: \leqslant \:\: 7 p_1^3 \: + \: 7^2 p_2^3 \: + \: \ldots
+ \: 7^{l - 2} p_{l - 2}^3 \\
& \qquad\qquad\qquad\qquad\qquad \:\:\:
+ \: 7^{l - 1} p_{l - 1}^3 \: + \: 7^{l - 1} p_l^3, \\
& p_1 \geqslant 0, \:\:\: p_2 \geqslant 0, \:\:\: \ldots, \:\:\: p_l \geqslant 0, \:\:\: l \, = \, 2, 3, 4, \ldots
\end{aligned}  \label{Eq_A_7}
\end{equation}

Let $ \bar{k} $ be small enough to satisfy
\begin{equation}
\begin{aligned}
& c_2 c_3 \bar{k} \, \leqslant \, \beta, \quad
c_2 c_3 \bar{k} \, \| \varphi \|^3_{\mathbf{C}^2 ([0, 2 \pi])} \: \leqslant \: \beta, \quad
0 < \beta < 1, \\
& 7^3 \beta^2 c_4^4 \, \leqslant \, 1, \quad 7^3 \beta^5 \, \leqslant \, 1.
\end{aligned}  \label{Eq_A_8}
\end{equation}
Then
\begin{equation}
\begin{aligned}
& \left\| u^1 \right\|_{\mathbf{C}^{2, 1} ([0, 2 \pi] \times [0, T])} \:\, \leqslant \:\,
c_4 \: + \: c_2 c_3 \bar{k} \, \left\| u^0 \right\|^3_{\mathbf{C}^{2, 1} ([0, 2 \pi] \times [0, T])} \\
& \qquad\qquad\qquad\quad \:\:\:
= \:\, c_4 \: + \: c_2 c_3 \bar{k} \, \| \varphi \|^3_{\mathbf{C}^2 ([0, 2 \pi])} \:\, \leqslant \:\,
c_4 + \beta, \\
& \left\| u^2 \right\|_{\mathbf{C}^{2, 1} ([0, 2 \pi] \times [0, T])} \:\, \leqslant \:\,
c_4 \: + \: c_2 c_3 \bar{k} \, \left\| u^1 \right\|^3_{\mathbf{C}^{2, 1} ([0, 2 \pi] \times [0, T])} \\
& \qquad\qquad\qquad \:
\leqslant \:\, c_4 \, + \, \beta (c_4 + \beta)^3  \:\, \leqslant \:\,
c_4 \, + \, 7 \beta \left( c_4^3 + \beta^3 \right) \\
& \qquad\qquad\qquad \:
\leqslant \:\, c_4 \, + \, 7 \beta c_4^3 \, + \, 7 \beta^4.
\end{aligned}  \label{Eq_A_9}
\end{equation}

Let us now verify the general relation
\begin{equation}
\begin{aligned}
& \left\| u^m \right\|_{\mathbf{C}^{2, 1} ([0, 2 \pi] \times [0, T])} \:\, \leqslant \:\,
c_4 \, \sum_{i = 0}^{m - 1} \left( 7 \beta c_4^2 \right)^i \: + \: \left( 7 \beta^4 \right)^{m - 1}, \\
& m \, = \, 2, 3, 4, \ldots
\end{aligned}  \label{Eq_A_10}
\end{equation}
by induction. Since (\ref{Eq_A_9}) serves as the basis of induction ($ m = 2 $), it remains to show that (\ref{Eq_A_10})
implies
\begin{equation}
\left\| u^{m + 1} \right\|_{\mathbf{C}^{2, 1} ([0, 2 \pi] \times [0, T])} \:\, \leqslant \:\,
c_4 \, \sum_{i = 0}^m \left( 7 \beta c_4^2 \right)^i \: + \: \left( 7 \beta^4 \right)^m.
\label{Eq_A_11}
\end{equation}
Using (\ref{Eq_A_6})--(\ref{Eq_A_8}) and (\ref{Eq_A_10}), one obtains
$$
\begin{aligned}
& \left\| u^{m + 1} \right\|_{\mathbf{C}^{2, 1} ([0, 2 \pi] \times [0, T])} \:\: \leqslant \:\:
c_4 \: + \: c_2 c_3 \bar{k} \, \| u^m \|^3_{\mathbf{C}^{2, 1} ([0, 2 \pi] \times [0, T])} \\
& \quad
\leqslant \:\: c_4 \:\, + \:\, \beta \, \left( \sum_{i = 0}^{m - 1} (7 \beta)^i \, c_4^{2i + 1}\: + \:
\left( 7 \beta^4 \right)^{m - 1} \right)^3 \\
& \quad
\leqslant \:\: c_4 \:\, + \:\, \beta \, \left( \sum_{i = 0}^{m - 1} 7^{i + 1} \, (7 \beta)^{3i} \, c_4^{3 (2i + 1)} \: + \:
7^m \left( 7 \beta^4 \right)^{3 (m - 1)} \right) \\
& \quad
= \:\: c_4 \:\, + \:\, c_4 \, \sum_{i = 1}^m \beta \, 7^i \, (7 \beta)^{3i - 3} \, c_4^{6i - 4} \:\, + \:\,
\beta \, 7^m \left( 7 \beta^4 \right)^{3m - 3} \\
& \quad
\leqslant \:\: c_4 \:\, + \:\, c_4 \, \sum_{i = 1}^m \left( 7 \beta c_4^2 \right)^i \,
\left( 7^{3i - 3} \, \beta^{2i - 2} \, c_4^{4i - 4} \right) \:\, + \:\,
\left( 7 \beta^4 \right)^m \, \left( 7^{3m - 3} \, \beta^{8m - 11} \right) \\
& \quad
\leqslant \:\: c_4 \:\, + \:\, c_4 \, \sum_{i = 1}^m \left( 7 \beta c_4^2 \right)^i \,
\left( 7^3 \beta^2 c_4^4 \right)^{i - 1} \:\, + \:\, \left( 7 \beta^4 \right)^m \, \left( 7^3 \beta^5 \right)^{m - 1} \,
\beta^{3m - 6}.
\end{aligned}
$$
This leads to (\ref{Eq_A_11}), because $ \, 7^3 \beta^2 c_4^4 \leqslant 1 $, $ \, 7^3 \beta^5 \leqslant 1, \, $ and
$ \, \beta^{3m - 6} \leqslant \beta^0 = 1 \, $ ($ m \geqslant 2 $) in line with (\ref{Eq_A_8}). Thus, (\ref{Eq_A_10}) holds,
and, consequently,
$$
\begin{aligned}
& \left\| u^m \right\|_{\mathbf{C}^{2, 1} ([0, 2 \pi] \times [0, T])} \:\, \leqslant \:\,
c_4 \, \sum_{i = 0}^{\infty} \left( 7 \beta c_4^2 \right)^i \: + \: 1 \:\, < \:\, \infty, \\
& m \, = \, 2, 3, 4, \ldots
\end{aligned}
$$
(one has $ \: 7 \beta c_4^2 \, = \, 7^{-\frac{1}{2}} \, (7^3 \beta^2 c_4^4)^{\frac{1}{2}} \, < \, 1 \: $ and
$ \, 7 \beta^4 \, = \, \left( 7^3 \beta^5 \right)^{\frac{1}{3}} \, \beta^{\frac{7}{3}} \, < \, 1 \, $ according to
(\ref{Eq_A_8})), which completes the proof.

\subsection{Proof of Theorem~\ref{Thm_3}}  \label{App_3}

By virtue of (\ref{Eq_11})--(\ref{Eq_13}), one has
$$
\begin{aligned}
\left\| u^{m + 1} - u^m \right\|_{\mathbf{C}^{2, 1} ([0, 2 \pi] \times [0, T])} \:\, = \:\,
\left\| \mathcal{E}_2 \left( \mathcal{F} u^m - \mathcal{F} u^{m - 1} \right)
\right\|_{\mathbf{C}^{2, 1} ([0, 2 \pi] \times [0, T])} & \\
\leqslant \:\, c_2 \, \left\| \mathcal{F} u^m - \mathcal{F} u^{m - 1}
\right\|_{\mathbf{C}^{2, 1} ([0, 2 \pi] \times [0, T])} \, , \quad
m \, = \, 1, 2, 3, \ldots &
\end{aligned}
$$
If one obtains the estimate
\begin{equation}
\begin{aligned}
& \left\| \mathcal{F} u^m - \mathcal{F} u^{m - 1} \right\|_{\mathbf{C}^{2, 1} ([0, 2 \pi] \times [0, T])} \\
& \qquad\qquad
\leqslant \:\, c_5 \bar{k} \, \left\| u^m - u^{m - 1} \right\|_{\mathbf{C}^{2, 1} ([0, 2 \pi] \times [0, T])} \, , \\
& m \, = \, 1, 2, 3, \ldots, \quad c_5 = \mathrm{const} > 0,
\end{aligned}  \label{Eq_A_12}
\end{equation}
then
$$
\begin{aligned}
& \left\| u^{m + 1} - u^m \right\|_{\mathbf{C}^{2, 1} ([0, 2 \pi] \times [0, T])} \:\, \leqslant \:\,
c_2 c_5 \bar{k} \, \left\| u^m - u^{m - 1} \right\|_{\mathbf{C}^{2, 1} ([0, 2 \pi] \times [0, T])} \, , \\
& m \, = \, 1, 2, 3, \ldots,
\end{aligned}
$$
and the Banach fixed point theorem (see, e.\,g., \cite[Theorem~4.16]{Muscat2014}) implies the sought-after result
(note that $ c_2 c_5 \bar{k} < 1 $ for sufficiently small $ \bar{k} $). But (\ref{Eq_A_12}) follows from
the representations
$$
\begin{aligned}
& \mathcal{F} u^m(x, t) \, - \, \mathcal{F} u^{m - 1}(x, t) \\
& \qquad
= \:\: u^m(x, t) \: (k(x) \, u^m(x - h, \, t) \: - \: f[u^m(\cdot, t)]) \\
& \qquad\quad \:\:
- \:\, u^{m - 1}(x, t) \, \left( k(x) \, u^{m - 1}(x - h, \, t) \: - \: f \left[ u^{m - 1}(\cdot, t) \right] \right) \\
& \qquad
= \:\: k(x) \, u^m(x - h, \, t) \, \left( u^m(x, t) \, - \, u^{m - 1}(x, t) \right) \\
& \qquad\quad \:\:
+ \:\, k(x) \, u^{m - 1}(x, t) \, \left( u^m(x - h, \, t) \, - \, u^{m - 1}(x - h, \, t) \right) \\
& \qquad\quad \:\:
- \:\, u^m(x, t) \, \left( f[u^m(\cdot, t)] \, - \, f \left[ u^{m - 1}(\cdot, t) \right] \right) \\
& \qquad\quad \:\:
- \:\, f \left[ u^{m - 1}(\cdot, t) \right] \, \left( u^m(x, t) \, - \, u^{m - 1}(x, t) \right), \\
\end{aligned}
$$
$$
\begin{aligned}
& f \left[ u^{m - 1}(\cdot, t) \right] \:\, = \:\,
\int_0^{2 \pi} k(x) \, u^{m - 1}(x, t) \, u^{m - 1}(x - h, \, t) \, \mathrm{d} x, \\
& f[u^m(\cdot, t)] \:\, = \:\, \int_0^{2 \pi} k(x) \, u^m(x, t) \, u^m(x - h, \, t) \, \mathrm{d} x,
\end{aligned}
$$
$$
\begin{aligned}
& f[u^m(\cdot, t)]) \: - \: f \left[ u^{m - 1}(\cdot, t) \right] \\
& \qquad
= \:\: \int_0^{2 \pi} k(x) \, u^m(x - h, \, t) \, \left( u^m(x, t) \, - \, u^{m - 1}(x, t) \right) \, \mathrm{d} x \\
& \qquad\quad \:\:
+ \:\, \int_0^{2 \pi} k(x) \, u^{m - 1}(x, t) \,
\left( u^m(x - h, \, t) \, - \, u^{m - 1}(x - h, \, t) \right) \, \mathrm{d} x,
\end{aligned}
$$
$ m = 1, 2, 3, \ldots, \, $ and the uniform boundedness of
$ \, \left\{ \| u^m \|_{\mathbf{C}^{2, 1} ([0, 2 \pi] \times [0, T])} \right\}_{m = 0}^{\infty} \, $
(the latter takes place for sufficiently small $ \bar{k} $ due to Lemma~\ref{Lemma_2}). This completes the proof.

\subsection{Proof of Theorem~\ref{Thm_4}}  \label{App_4}

For $ h \in (0, h_0] $, let $ x^*_h \in [0, 2 \pi] $ and $ t^*_h \in [0, T] $ satisfy
\begin{equation}
u_h(x^*_h, t^*_h) \: = \: \min_{ \substack{0 \leqslant x \leqslant 2 \pi, \\ 0 \leqslant t \leqslant T} } u_h(x, t).
\label{Eq_A_13}
\end{equation}
The first-order necessary minimum conditions imply
\begin{equation}
\begin{aligned}
& \frac{\partial u_h}{\partial t} \, (x^*_h, t^*_h) \: \geqslant \: 0 \quad \mathrm{for} \:\:\, t^*_h = 0, \\
& \frac{\partial u_h}{\partial t} \, (x^*_h, t^*_h) \: = \: 0 \quad \mathrm{for} \:\:\, 0 < t^*_h < T, \\
& \frac{\partial u_h}{\partial t} \, (x^*_h, t^*_h) \: \leqslant \: 0 \quad \mathrm{for} \:\:\, t^*_h = T.
\end{aligned}  \label{Eq_A_14}
\end{equation}
If $ \, 0 < x^*_h < 2 \pi, \, $ the first- and second-order necessary minimum
conditions yield $ \: \dfrac{\partial u_h}{\partial x} \, (x^*_h, t^*_h) \, = \, 0 \: $ and
\begin{equation}
\frac{\partial^2 u_h}{\partial x^2} \, (x^*_h, t^*_h) \: \geqslant \: 0.  \label{Eq_A_15}
\end{equation}
If $ x^*_h = 0 $ or $ x^*_h = 2 \pi $, then
$$
\begin{aligned}
& u_h(x^*_h, t^*_h) \: = \: u_h(0, t^*_h) \: = \: u_h(2 \pi, t^*_h), \\
& \frac{\partial u_h}{\partial x} \, (x^*_h, t^*_h) \: = \: \frac{\partial u_h}{\partial x} \, (0, t^*_h) \: = \:
\frac{\partial u_h}{\partial x} \, (2 \pi, t^*_h)
\end{aligned}
$$
(by virtue of the boundary conditions in (\ref{Eq_6})), and the minimum can be reached at both points
$ (0, t^*_h) $, $ (2 \pi, t^*_h) $ only for $ \: \dfrac{\partial u_h}{\partial x} \, (x^*_h, t^*_h) \, = \, 0, \: $
so Taylor's theorem with the Lagrange remainder again leads to (\ref{Eq_A_15}).

Since $ \: u_h, \, \partial u_h / \partial x, \, \partial^2 u_h / \partial x^2 \: $ are continuous on
$ \, [0, 2 \pi] \times [0, T] \, $ and $ \: u_h(x - h, \, t) \, = \, u_h(x - h + 2 \pi, \, t) \: $ for
$ \, -h \leqslant x - h \leqslant 0, \, $ Taylor's theorem with Lagrange remainder and the condition~(\ref{Eq_17}) imply
\begin{equation}
\begin{aligned}
& \max_{ \substack{0 \leqslant x \leqslant 2 \pi, \\ 0 \leqslant t \leqslant T} } \,
\left| u_h(x - h, \, t) \: - \: \left( u_h(x, t) \: - \: h \, \frac{\partial u_h(x, t)}{\partial x} \right) \right|
\:\: \leqslant \:\: c_6 h^2 \\
& \forall h \in (0, h_0],
\end{aligned}  \label{Eq_A_16}
\end{equation}
where
\begin{equation}
\begin{aligned}
c_6 \:\, = \:\, \sup_{0 < h \leqslant h_0} \,
\max_{ \substack{0 \leqslant x \leqslant 2 \pi, \\ 0 \leqslant t \leqslant T} } \,
\left| \frac{\partial^2 u_h(x, t)}{\partial x^2} \right| \:\, < \:\, \infty.
\end{aligned}  \label{Eq_A_17}
\end{equation}
Hence,
$$
\begin{aligned}
& f[u_h(\cdot, t)]) \:\, = \:\, \int_0^{2 \pi} k(x) \, u_h(x, t) \, u_h(x - h, \, t) \, \mathrm{d} x \\
& \quad
\:\, = \:\, \int_0^{2 \pi} k(x) \, (u_h(x, t))^2 \, \mathrm{d} x \: - \: h \,
\int_0^{2 \pi} k(x) \, u_h(x, t) \, \frac{\partial u_h(x, t)}{\partial x} \, \mathrm{d} x \: + \: o(h, t) \\
& \forall h \in (0, h_0],
\end{aligned}
$$
where
\begin{equation}
\lim_{h \, \to \, +0} \, \frac{\max\limits_{0 \leqslant t \leqslant T} \, |o(h, t)|}{h} \:\, = \:\, 0.  \label{Eq_A_18}
\end{equation}
Using integration by parts and the boundary conditions for $ k $ and $ u_h $ (see Assumption~\ref{Assum_1} and (\ref{Eq_6})),
one arrives at
$$
\begin{aligned}
& \int_0^{2 \pi} k(x) \, u_h(x, t) \, \frac{\partial u_h(x, t)}{\partial x} \, \mathrm{d} x \\
& \qquad
= \:\, \frac{1}{2} \left. \left( k(x) \, (u_h(x, t))^2 \right) \right|_{x = 0}^{x = 2 \pi} \: - \:
\frac{1}{2} \, \int_0^{2 \pi} k'(x) \, (u_h(x, t))^2 \, \mathrm{d} x \\
& \qquad
= \:\, -\frac{1}{2} \, \int_0^{2 \pi} k'(x) \, (u_h(x, t))^2 \, \mathrm{d} x \\
& \forall h \in (0, h_0],
\end{aligned}
$$
and, therefore,
$$
f[u_h(\cdot, t)]) \:\, = \:\, \int_0^{2 \pi} \left( k(x) \, + \, \frac{h}{2} \, k'(x) \right) \,
(u_h(x, t))^2 \, \mathrm{d} x \: + \: o(h, t) \quad \forall h \in (0, h_0].
$$
According to Assumption~\ref{Assum_1}, $ k $ is continuously differentiable and positive on $ [0, 2 \pi] $, so there exist
constants $ c_7 > 0 $ and $ h'_0 \in (0, h_0] $ such that
$$
\min_{0 \leqslant x \leqslant 2 \pi} \, \left\{ k(x) \, + \, \frac{h}{2} \, k'(x) \right\} \:\, \geqslant \:\, c_7 \quad
\forall h \in (0, h'_0],
$$
leading to
$$
f[u_h(\cdot, t)]) \:\, \geqslant \:\, c_7 \int_0^{2 \pi} (u_h(x, t))^2 \, \mathrm{d} x \: + \: o(h, t) \quad
\forall h \in (0, h'_0].
$$
Furthermore, the normalization condition~(\ref{Eq_4}) and the Cauchy--Schwarz inequality for the Hilbert space
$ \, \mathcal{L}^2 ([0, 2 \pi]; \, \mathbb{R}) \, $ yield
$$
1 \:\, = \:\, \left( \int_0^{2 \pi} 1 \cdot u_h(x, t) \, \mathrm{d} x \right)^2 \:\, \leqslant \:\,
2 \pi \int_0^{2 \pi} (u_h(x, t))^2 \, \mathrm{d} x
$$
and, consequently,
$$
f[u_h(\cdot, t)]) \: \geqslant \: \frac{c_7}{2 \pi} \, + \, o(h, t) \quad \forall h \in (0, h'_0].
$$
Together with (\ref{Eq_A_18}), this implies the existence of a constant $ h''_0 \in (0, h'_0] $ such that
\begin{equation}
c_8 \:\, = \:\, \inf_{0 < h \leqslant h''_0} \, \min\limits_{0 \leqslant t \leqslant T} \,
f[u_h(\cdot, t)]) \:\, > \:\, 0.  \label{Eq_A_19}
\end{equation}

Let also
\begin{equation}
w_h(x, t) \: = \: u_h(x, t) \, e^{\Phi_h(t)}, \quad \Phi_h(t) \: = \: \int_0^t f[u_h(\cdot, s)] \, \mathrm{d} s \quad
\forall h \in (0, h_0].  \label{Eq_A_20}
\end{equation}
Then
\begin{equation}
\begin{aligned}
& \frac{\partial w_h}{\partial t} \, (x^*_h, t^*_h) \:\: = \:\:
\frac{\partial u_h}{\partial t} \, (x^*_h, t^*_h) \: e^{\Phi_h(t^*_h)} \\
& \qquad
+ \:\, u_h(x^*_h, t^*_h) \: e^{\Phi_h(t^*_h)} \: f[u_h(\cdot, t^*_h)] \quad \forall h \in (0, h_0].
\end{aligned}  \label{Eq_A_21}
\end{equation}
Moreover, (\ref{Eq_6}) and (\ref{Eq_A_20}) lead to
$$
\begin{aligned}
\frac{\partial w_h}{\partial t} \, (x^*_h, t^*_h) \:\: = \:\: k(x^*_h) \: u_h(x^*_h, t^*_h) \: u_h(x^*_h - h, \, t^*_h) \:
e^{\Phi_h(t^*_h)} & \\
+ \:\, \alpha \: \frac{\partial^2 u_h}{\partial x^2} \, (x^*_h, t^*_h) \: e^{\Phi_h(t^*_h)} \quad
\forall h \in (0, h_0], &
\end{aligned}
$$
and, therefore,
$$
\begin{aligned}
& \frac{\partial w_h}{\partial t} \, (x^*_h, t^*_h) \:\: \geqslant \:\: k(x^*_h) \: u_h(x^*_h, t^*_h) \:
u_h(x^*_h - h, \, t^*_h) \: e^{\Phi_h(t^*_h)} \\
& \forall h \in (0, h_0]
\end{aligned}
$$
due to (\ref{Eq_A_15}) and $ \alpha > 0 $. Together with the relations~(\ref{Eq_17}), (\ref{Eq_A_16}) and
the continuity of $ k $ on $ [0, 2 \pi] $, this yields the existence of a constant~$ c_9 > 0 $ such that
\begin{equation}
\begin{aligned}
& \frac{\partial w_h}{\partial t} \, (x^*_h, t^*_h) \:\: \geqslant \:\: k(x^*_h) \: (u_h(x^*_h, t^*_h))^2 \:
e^{\Phi_h(t^*_h)} \\
& \qquad\qquad\qquad\quad \:\,
- \:\, c_9 h \: |u_h(x^*_h, t^*_h)| \: e^{\Phi_h(t^*_h)} \quad \forall h \in (0, h_0].
\end{aligned}  \label{Eq_A_22}
\end{equation}

Let us now establish the sought-after property by contradiction. Assume the existence of a sequence
$ \, \{ h_i \}_{i = 1}^{\infty} \, \subset \, (0, h''_0] \, $ such that $ \: \lim\limits_{i \to \infty} h_i \, = \, 0 \: $
and $ \: u_{h_i}(x^*_{h_i}, t^*_{h_i}) \, < \, 0 \: $ for all $ i \in \mathbb{N} $. Then $ t^*_{h_i} \neq 0 $ (since
$ \, u_{h_i}(x, 0) = \varphi(x) \geqslant 0 $), (\ref{Eq_A_14}) reduces to
$ \: \dfrac{\partial u_{h_i}}{\partial t} \, (x^*_{h_i}, t^*_{h_i}) \, \leqslant \, 0, \: $ and (\ref{Eq_A_21}) leads to
\begin{equation}
\begin{aligned}
\frac{\partial w_{h_i}}{\partial t} \, (x^*_{h_i}, t^*_{h_i}) \:\: \leqslant \:\:
u_h(x^*_{h_i}, t^*_{h_i}) \: e^{\Phi_{h_i}(t^*_{h_i})} \: f[u_{h_i}(\cdot, t^*_{h_i})].
\end{aligned}  \label{Eq_A_23}
\end{equation}
From (\ref{Eq_A_22}) and (\ref{Eq_A_23}), one obtains
$$
\begin{aligned}
u_{h_i}(x^*_{h_i}, t^*_{h_i}) \: f[u_{h_i}(\cdot, t^*_{h_i})] \:\: \geqslant \: & \:
k(x^*_{h_i}) \: (u_{h_i}(x^*_{h_i}, t^*_{h_i}))^2 \\
& - \:\, c_9 h_i \: |u_{h_i}(x^*_{h_i}, t^*_{h_i})|,
\end{aligned}
$$
and, consequently,
$$
f[u_{h_i}(\cdot, t^*_{h_i})] \:\: \leqslant \:\: k(x^*_{h_i}) \: u_{h_i} (x^*_{h_i}, t^*_{h_i}) \:\, + \:\, c_9 h_i
$$
(due to $ \: u_{h_i}(x^*_{h_i}, t^*_{h_i}) \, < \, 0 $). Together with (\ref{Eq_A_19}) and the positivity of
$ k $ on $ [0, 2 \pi] $, this implies that $ \, 0 < c_8 < c_9 h_i \, $ for all $ i \in \mathbb{N} $, which
contradicts with $ \: \lim\limits_{i \to \infty} h_i \, = \, 0 $. Thus, $ u_h $ should indeed be a nonnegative function
for sufficiently small~$ h $.

\end{document}